\documentclass[sigconf]{acmart}
\setcopyright{none}
\acmConference[ASE 2020]{The 35th IEEE/ACM International Conference on Automated Software Engineering}{21 - 25 September}{Melbourne, Australia}

\def\BibTeX{{\rm B\kern-.05em{\sc i\kern-.025em b}\kern-.08emT\kern-.1667em\lower.7ex\hbox{E}\kern-.125emX}}

\usepackage{graphicx}
\usepackage{array}
\usepackage{url}
\usepackage{fancybox}
\usepackage{multirow}
\usepackage{color}
\usepackage{colortbl}
\usepackage{balance}
\usepackage{fancyhdr}
\usepackage{subfig}
\usepackage{diagbox}
\usepackage{bbding}
\usepackage{placeins}
\usepackage{booktabs}
\usepackage{enumitem}
\usepackage{xcolor,lipsum}

\usepackage{amsmath,amssymb,amsfonts}
\usepackage{textcomp}
\usepackage{algorithmic}
\usepackage{listings}
\usepackage{color}
\usepackage{commath}
\usepackage{comment}

\definecolor{light-gray}{rgb}{.906,  .902,  .902}

\usepackage{bm}

\begin{document}
% Title portion. Note the short title for running heads
\title{When Deep Learning Meets Smart Contracts}

\begin{abstract}
Ethereum has become a widely used platform to enable secure, Blockchain-based financial and business transactions. However, many identified bugs and vulnerabilities in smart contracts have led to serious financial losses, which raises serious concerns about smart contract security. Thus, there is a significant need to better maintain smart contract code and ensure its high reliability.

In this research: (1) Firstly, we propose an automated deep learning based approach to learn structural code embeddings of smart contracts in Solidity, which is useful for clone detection, bug detection and contract validation on smart contracts. We apply our approach to more than 22K solidity contracts collected from the Ethereum blockchain, results show that the clone ratio of solidity code is at around 90\%, much higher than traditional software. % Our work reveals homogeneous of the Ethereum ecosystem. 
We collect a list of 52 known buggy smart contracts belonging to 10 kinds of common vulnerabilities as our bug database.
Our approach can identify more than 1000 clone related bugs based on our bug databases efficiently and accurately. 
(2) Secondly, according to developers' feedback, we have implemented the approach in a web-based tool, named {\sc SmartEmbed}, to facilitate Solidity developers for using our approach. 
Our tool can assist Solidity developers to efficiently identify repetitive smart contracts in the existing Ethereum blockchain, as well as checking their contract against a known set of bugs, which can help to improve the users' confidence in the reliability of the contract.
We optimize the implementations of {\sc SmartEmbed} which is sufficient in supporting developers in real-time for practical uses. 
The Ethereum ecosystem as well as the individual Solidity developer can both benefit from our research.

{\sc SmartEmbed} website: \url{http://www.smartembed.tools}

Demo video: 
\url{https://youtu.be/o9ylyOpYFq8}

Replication package:
\url{https://github.com/beyondacm/SmartEmbed}
\end{abstract}

% <concept_id>10011007.10011074.10011111.10011113</concept_id>
% <concept_desc>Software and its engineering~Software evolution</concept_desc>
% <concept_id>10011007.10011074.10011111.10011696</concept_id>
% <concept_desc>Software and its engineering~Maintaining software</concept_desc>

% \ccsdesc[500]{Software and its engineering~Software evolution}
% \ccsdesc[500]{Software and its engineering~Maintaining software}
\author{Zhipeng Gao}
\affiliation{%
  \institution{Monash University, Australia}
  }
% \author{\hspace{-6mm} Vinoj Jayasundara, Lingxiao Jiang}
% \affiliation{%
%   \institution{\hspace{-6mm} Singapore Management University, Singapore}
%   }
% \author{Xin Xia}
% \affiliation{%
%   \institution{Monash University, Australia}
%   }  
%   \author{David Lo}
% \affiliation{%
%   \institution{Singapore Management University, Singapore}
%   }
% \author{John Grundy}
% \affiliation{%
%   \institution{Monash University, Australia}
%   } 

%\keywords{Smart Contract, Clone Detection, Bug Detection, Code Embedding}

\renewcommand{\shortauthors}{Gao et al.}

\maketitle

\section{Introduction}
\label{sec:intro}
In recent years, along with the adoption and development of cryptocurrencies on distributed ledgers (a.k.a., blockchains), \emph{smart contracts}~\cite{Szabo1994} has attracted more and more attention with Ethereum blockchain platform. 
A Smart contract is a computer program that can be triggered to execute any task when specifically predefined conditions are satisfied. A major concern in the Ethereum platform is the security of smart contracts. A smart contract in the blockchain often involves cryptocurrencies worthy of millions of USD (e.g., DAO\footnote{\url{https://en.wikipedia.org/wiki/TheDAO(organization)}}, Parity\footnote{\url{https://paritytech.io/security-alert-2/}} and many more). 
Moreover, different from a traditional software program, the smart contract code is immutable after its deployment. They cannot be changed but may be killed when any security issue is identified within the smart contracts. 
This introduces challenges to blockchain maintenance and gives much incentive to hackers for discovering and exploiting potential problems in smart contracts, and there is a very significant need to check and ensure the robustness of smart contracts before deployment.

Many prior works have investigated characteristics of bugs in smart contracts and underlying blockchain systems (e.g., \cite{Wan2017, li2017survey, atzei2017survey, bartoletti2017empirical, chen2017under}) and detection of smart contract bugs (e.g., \cite{Bhargavan2016, brownformal, luu2016making, tsankov2018securify, tikhomirov2018smartcheck, delmolino2016step, Mueller2018}). A major disadvantage is that all these existing tools require certain bug patterns or specification rules defined by human experts. Considering the high stakes in smart contracts and race between attackers and defenders, it can be far too slow and costly to write new rules and construct new checkers in response to new bugs and exploits created by attackers.
Recently, there are also studies on clones and clone detection for Ethereum smart contracts (e.g., \cite{He2019,Eclone2018}). However, they use expensive symbolic transaction sketches or pairwise comparisons which affect their efficiency and they are limited to clone detection.

In this research, we aim to address these problems by exploring deep learning techniques to learn vector representation for smart contracts.
Our main idea is two folds: 
(1) Structural Code Embeddings: code and bug patterns, including their lexical, syntactical, and even some semantic information, can be automatically encoded into numerical vectors via techniques adapted from deep learning and word embeddings (e.g., \cite{ye2016word, bojanowski2016enriching, mikolov2013efficient, mikolov2013distributed, turian2010word}). 
(2) Similarity Checking: code checking can be essentially done through similarity checking among the numerical vectors representing various kinds of code elements of various levels of granularity in smart contracts.
% Different tasks such as code clone detection, bug detection, and code validation can be viewed as variants of the problem of finding ``similar'' code. 

In our previous work~\cite{gao2020checking}, we proposed an deep learning based approach that can efficiently and effectively check smart contracts with structural code embeddings. 
With the help of suitable concrete code embedding and similarity checking techniques, our approach can be general enough to be applied for various code debugging and maintenance tasks.
These include repetitive (a.k.a. duplicate or cloned) contract detection, detection of specific kinds of bugs in a large contract corpus, or validation of a contract against a set of known bugs. 
Moreover, our approach can easily add new bug checking rules by generating code embeddings for the changing bug patterns, without extra manual efforts in defining the bug specifications. 
According to the feedback of Solidity developers in Github, we further implemented our approach as a web-based tool, named {\sc SmartEmbed}, which can help Solidity developers check code clones and bugs for their own smart contracts~\cite{gao2019smartembed}. 
Furthermore, to meet the efficiency requirements as an online web tool, we improve the efficiency of {\sc SmartEmbed} in three ways: (i) replace multiple loop structure calculation with matrix computation. (ii) put the code embeddings into cache to reduce redundant data loading. (iii) create indexes for smart contracts in our database to speed up information retrieval process. 
Considering the rapid increase in the number of smart contracts, we automatically updated our online model with newly added blocks from Ethereum blockchain, so developers can keep up to date with new changes by using our tool.

\begin{figure}\vspace{-0.0cm}
\centerline{\includegraphics[width=0.50\textwidth]{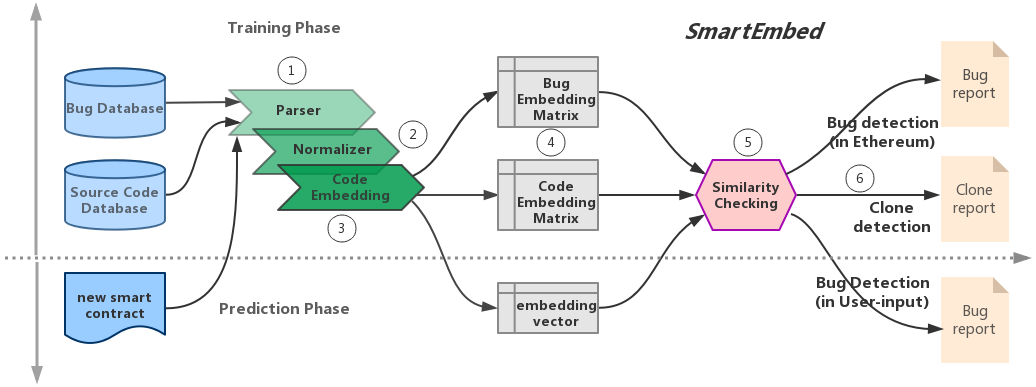}}
 \vspace*{-20pt}
\caption{Overview of Our Approach}
\label{fig:approach}
 \vspace{-0.4cm}
\end{figure}

The rest of the paper is organized as follows.
Section~\ref{sec:related} discusses related work. 
Section~\ref{sec:approach} presents the overall workflow of our approach and details of each step. 
Section~\ref{sec:eval} shows the experimental results.
Section~\ref{sec:con} concludes the contribution of our work.

\section{Related Work}\label{sec:related}

Both clone detection and bug detection have been extensively studied in the literature for both traditional software programs and Ethereum smart contracts. Clone detection techniques utilize expensive symbolic transaction sketch~\cite{Eclone2018} or pair-wise signature comparison~\cite{He2019}. They can be applied to compiled contract bytecode, but efficiency is limited for checking contract-level similarities, instead of more fine-grained levels of granularity. More bug detection techniques have been developed for smart contracts (e.g., \cite{Bhargavan2016, brownformal, luu2016making, tsankov2018securify, tikhomirov2018smartcheck, delmolino2016step, Mueller2018}), while they often require manual efforts in defining specifications and/or bug patterns needed for validation and/or bug checking.

Machine learning and deep learning techniques have been used for clone detection (e.g., \cite{astnn2019,cclearner2017,White2016DeepLC}) and bug detection problems (e.g. \cite{Yang2015,Li2017}) in traditional software programs too, but little has been applied for smart contracts. Our approach is unique in that it utilizes deep learning and similarity checking techniques to unify clone detection and bug detection together efficiently and accurately for Ethereum smart contracts.

\section{Approach}
\label{sec:approach}
%New colors defined below
\definecolor{codegreen}{rgb}{0,0.6,0}
\definecolor{codegray}{rgb}{0.5,0.5,0.5}
\definecolor{codepurple}{rgb}{0.58,0,0.82}
\definecolor{backcolour}{rgb}{0.96,0.96,0.96}

%Code listing style named "mystyle"
\lstdefinestyle{mystyle}{
  backgroundcolor=\color{backcolour},   commentstyle=\color{codegreen},
  keywordstyle=\color{magenta},
  numberstyle=\tiny\color{codegray},
  stringstyle=\color{codepurple},
  basicstyle=\footnotesize,
  breakatwhitespace=false,         
  breaklines=true,                 
  captionpos=b,                    
  keepspaces=true,                 
  numbers=left,                    
  numbersep=5pt,                  
  showspaces=false,                
  showstringspaces=false,
  showtabs=false,                  
  tabsize=2
}
\lstset{style=mystyle}
\title{SmartEmbed}
%\section{Code examples}
%Python code highlighting

% \subsection{Overview}\label{AA}
Fig.\ref{fig:approach} illustrates the overall framework of Our approach. Based on code embeddings and similarity checking, our approach targets two tasks in a unified approach: clone detection and bug detection.
For clone detection, our approach can identify similar smart contracts. 
For bug detection, based on our bug database, our approach can detect bugs in the existing contracts in the Ethereum blockchain and/or in any smart contract given by solidity developers that are similar to any known bug in the bug database.
Our approach contains two phases: a model training phase and a prediction phase.

There are mainly 4 steps during the model training phase. We built a custom Solidity parser for smart contract source code. The parser generates an abstract syntax tree (ASTs) for each smart contract in our collected dataset, and serializes the parse tree into a stream of tokens depending on the types of the tree nodes (step 1). After that, the normalizer reassembles the token streams to eliminate nonessential differences (e.g., the stop word, values of constants or literals) between smart contracts (step 2). The output token streams are then fed into our code embedding learning module, and each code fragment is embedded into a fixed-dimension numerical vector (step 3). After the code embedding learning step, all the source code is encoded into the source code embedding matrix; in the meanwhile, all the bug statements we collected are encoded into the bug embedding matrix (step 4).

In the prediction phase, any given new smart contract is turned into embedding vectors by going through the steps 1,2,3 and utilizing the learned embedding matrices. Similarity comparison is performed between the embeddings for the given contract and those in our collected database (step 5), and similarity thresholds are used to govern whether a code fragment in the given contract will be considered as code clones or clone-related bugs (step 5-6).

% \section{Implementation Details \& Tool Usage}
% \label{sec:implem}
% \input{implem}

\section{Evaluation Results}
\label{sec:eval}
% We compared {\sc SmartEmbed} with two well-known tools that are specific for clone detection (DECKARD~\cite{jiang2007deckard} extended for Solidity) and bug detection (SmartCheck~\cite{tikhomirov2018smartcheck}) respectively.

\smallskip
\noindent{\itshape\bfseries Clone Detection}.
Our approach can effectively identify many repetitive Solidity codes in Ethereum blockchain. 
Our experimental results against more than 22K smart contracts show the clone ratio of solidity code is at around 90\%, much higher than traditional software, which reveals homogeneous of the Ethereum ecosystem. Our approach can detect more semantic clones accurately than the commonly used clone detection tool Deckard~\cite{jiang2007deckard}.
The relatively high ratio of code clones in smart contracts may cause severe threats, such as security attacks, resource wastage, etc. Finding such clones can enable significant applications such as vulnerability discovery and deployment optimization (reduce contract size and duplication), hence contributing to the overall health of the Ethereum ecosystem.
Our work in identifying clones can also help Solidity developers to check for plagiarism in smart contracts,
which may cause a huge financial loss to the original contract creator.

\smallskip
\noindent{\itshape\bfseries Bug Detection}.
For bug detection, we first collect a list of 52 known buggy smart contracts belonging to 10 kinds of common vulnerabilities.
The bug detection results show that our approach can efficiently and accurately identify more than 1,000 clone-related bugs based on our bug database in Ethereum blockchain, which can enable efficient checking of smart contracts with changing code and bug patterns. For contract validation, our approach can capture bugs similar to known ones with low false positive rates, the query for a clone or a bug is quite efficient which can be sufficient for practical uses. 
Our web-based tool can be easily updated with the newly added contracts and bug patterns in Ethereum blockchain and further support developers for using our approach.

\section{Summary and Contributions}
\label{sec:con}
This paper presented a deep learning based approach, {\sc SmartEmbed}, for detecting code clones and bugs in smart contracts accurately and efficiently.
It develops a code embedding technique for tokens and syntactical structures in Solidity code and utilizes similarity checking to search for ``similar'' code satisfying certain thresholds. Our approach can be easily updated to recognize new contract clones and new kinds of bugs when the contract code and bugs evolve.
The approach is automated on the contract and bug data collected from the Ethereum blockchain.
It helps developers to find repetitive contract code and clone-related bugs in existing contracts, which helps to improve developers' confidence in the reliability of their contracts.
It also helps to efficiently validate given smart contracts against known set of bugs without the need of manually defined bug patterns.
The clone detection and bug detection results can benefit the smart contract community as well as individual Solidity developers.

\balance
\bibliographystyle{ACM-Reference-Format}
\bibliography{samples}

\end{document}